\documentclass[prl,twocolumn,superscriptaddress,showpacs,floats,floatfix]{revtex4}
\usepackage{epsfig,dcolumn,amsmath,latexsym}
\begin{document}

\title{Formation of a Metallic Contact: Jump to Contact Revisited}

\preprint{1}

\author{C.\ Untiedt}
\thanks{Electronic mail address: untiedt@ua.es}
\affiliation{LT-NanoLab, Departamento de F\'\i sica Aplicada, Universidad
de Alicante, Campus de San Vicente del Raspeig, E-03690 Alicante, Spain.}

\author{M.J. Caturla}
\affiliation{LT-NanoLab, Departamento de F\'\i sica Aplicada, Universidad
de Alicante, Campus de San Vicente del Raspeig, E-03690 Alicante, Spain.}

\author{M.R. Calvo}
\affiliation{LT-NanoLab, Departamento de F\'\i sica Aplicada, Universidad
de Alicante, Campus de San Vicente del Raspeig, E-03690 Alicante, Spain.}

\author{J.\ J.\ Palacios}
\affiliation{LT-NanoLab, Departamento de F\'\i sica Aplicada, Universidad
de Alicante, Campus de San Vicente del Raspeig, E-03690 Alicante, Spain.}

\author{R.\,C.\ Segers }
\author{J.\,M.\ van Ruitenbeek}
\affiliation{Kamerlingh Onnes Laboratorium, Leiden University, PO Box
9504, NL-2300 RA Leiden, The Netherlands}

\date{\today}

\begin{abstract}
The transition from tunneling to metallic contact between two
surfaces does not always involve a jump, but can be smooth. We
have observed that the configuration and material composition of
the electrodes before contact largely determines the presence or
absence of a jump. Moreover, when jumps are found preferential
values of conductance have been identified. Through combination of
experiments, molecular dynamics, and first-principles transport
calculations these conductance values are identified with atomic
contacts of either monomers, dimers or double-bond contacts.
\end{abstract}


\maketitle
\begin{figure}[!b]
\epsfig{width=8cm,figure=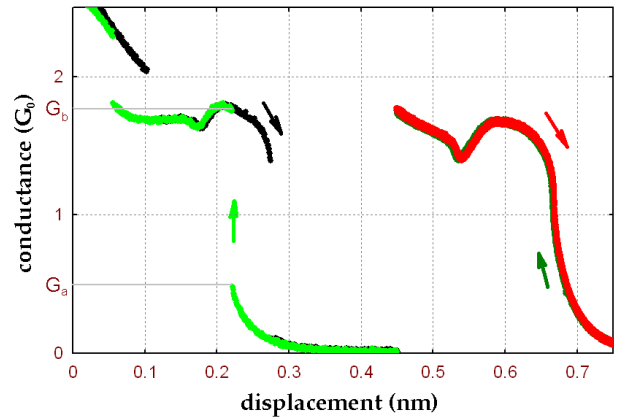} \caption{Traces of
conductance for Ir atomic contacts at 4.2K showing a jump to
contact (left) or a smooth transition (right). Arrows indicate the
direction of the displacement of the electrodes. When there is a
jump to contact, the jump starts from G$_a$ and ends at
G$_b$.}\label{Traces}
\end{figure}
\begin{figure}[!t]
\epsfig{width=8.5cm,figure=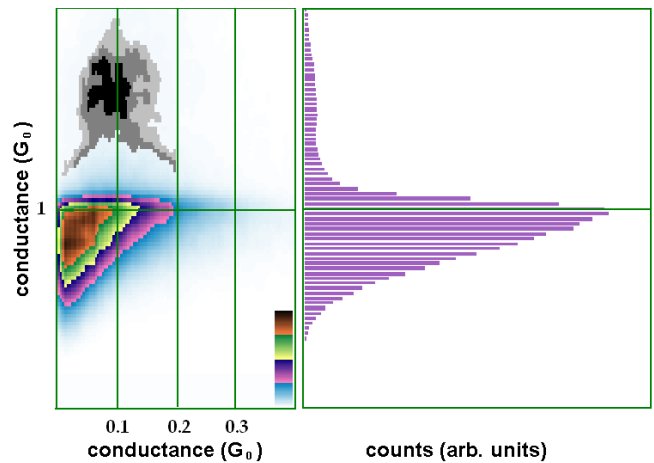} \caption{(color
online)Analysis of the steepest jump of conductance before the
formation of a metallic contact for the case of gold, made from
more than 300\,000 conductance traces. The left panel shows a
density plot, where the horizontal axes represents the conductance
at which the jump takes place and the vertical axes shows the
conductance of the contact formed. We have artificially changed
the colors of the peak above (gray scale) to make it visible. The
right panel shows the corresponding histogram of the conductance
of the contact formed after the jump.}\label{Density_Au}
\end{figure}
\begin{figure}[!b]
\epsfig{width=8.5cm,figure=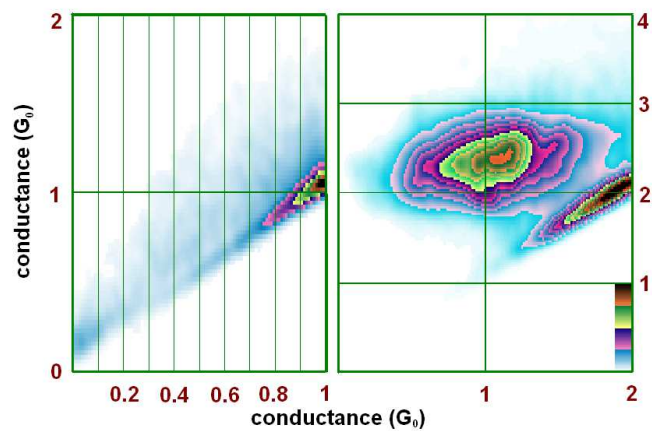}\\
\caption{(color online)Analysis of the steepest jump of
conductance before the formation of a metallic contact for the
case of W (left), and Ir (right) made from more than 10\,000 and
70\,000 conductance traces, respectively. The horizontal axes
represents the conductance at which the jump takes place and the
vertical axes shows the conductance of the contact
formed.}\label{Density_W_Ir}
\end{figure}
\begin{figure}[hb]
\epsfig{width=9cm,figure=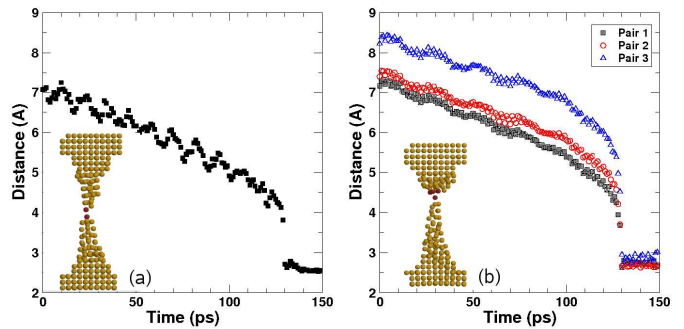} \caption{Distances between the
pairs of atoms that first come into contact during the approach for two
different configurations for Au obtained from molecular dynamics
simulations. The most common configuration (a) consists of a dimer, while
(b) shows the case of a monomer. The insets show the positions of the
atoms. Dark atoms are those at contact. }\label{Simulation01}
\end{figure}
\begin{figure}[ht]
\epsfig{width=5cm,figure=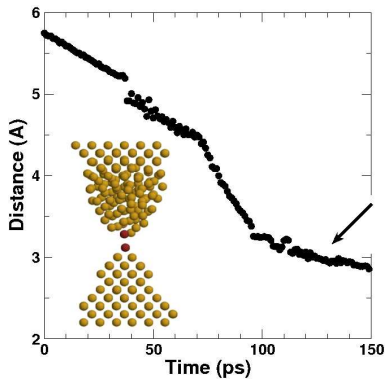} \caption{ Distances
between the two atoms that first come into contact for W, as
obtained from a molecular dynamics simulation. The inset shows the
positions of the atoms. Dark spheres are those atoms at contact.
Contact (defined with the cut-off between first and second nearest
neighbors) occurs at 130ps. Notice the different scale with
respect to Fig.~\protect\ref{Simulation01} }\label{Simulation02}
\end{figure}

Matter at the atomic scale has attracted much attention during
the last two decades. Not only because of the new properties arising when
size is decreased, but also because the mechanisms found there can help us
understand behavior at the macroscopic scale. This is the case of atomic
contacts where many new and unexpected phenomena have been
found\cite{Review} and, at the same time, have provided clues about many
technological problems in the macroscopic scale. A clear example would be
the study of
%phenomena such as
adhesion or friction which are consequences of the  formation of a contact
between two bodies, a process which always involves the formation of at
least one atomic contact\cite{Landman}.
\\
The formation of a contact between two metallic electrodes can be studied
by measuring the electronic transport through them. It has been observed
that right before the electrodes are brought into contact, electrons can
tunnel from one electrode to the other. This tunneling current, as shown
by the curve on the left in Fig. \ref{Traces}, will increase exponentially
while the electrodes come together and at some point a Jump to Contact
(JC) occurs: the electrode separation becomes unstable and a sudden
increase of the current takes place because of the formation of a
single-atom contact\cite{First,Review} with a conductance of about a
quantum unit(G$_0 = 2e^2/h$). When the electrodes are then separated, a
hysteretic behavior is observed in which the contact breaks in a jump to
form a vacuum gap while relieving the elastic tension accumulated
\cite{Gabino}. \\
The process described above has been seen as the general rule in
the formation of a metallic contact with a few exceptions to be
understood\cite{Review}. It has been reported that in some cases
the jump does not occur and, instead, there is a continuous
increase of current from tunneling to contact, after which it
saturates forming a plateau showing no hysteresis when the
electrodes are separated. Examples of this behavior have
accumulated and presently have been reported for the cases of a
W(111)-Au(111) contact\cite{Cross}, W nanocontacts\cite{tungsten},
and for the case when a STM tip approaches Ag and Cu adatoms on
(111) surfaces\cite{Adatoms}.
\\
Here, the formation of a metallic contact for various metals (Au,
Pt, Ag, Ir, W, and Ni) is investigated. A statistical analysis
using density plots has been developed that, combined with
atomistic simulations and ab-initio calculations of the electronic
transport, provide new insight into the problem. The absence of a
JC is not an exception but it occurs for many metals, in
particular for Ni, W and Ir.
\\
In order to control the formation of the atomic contacts we have
employed a home-made high stability Scanning Tunneling Microscope
(STM) and Mechanically Controlled Break Junctions\cite{Review}
(MCBJ). In STM experiments, the tip of the STM is contacted to a
sample made of the same material. The MCBJ experiments start from
a notched wire that is glued to a bending substrate which bends
using a piezo-element. At some point the wire breaks at the notch
forming two fresh surfaces which are brought back together to form
the atomic contacts. All experiments were performed at low
temperatures (4.2K) and under cryogenic vacuum conditions. The
traces of conductance were taken at constant bias voltage
(typically at 10mV) which has been shown\cite{Sirvent} to have
negligible effect on the JC. A current amplifier in the range of
mA was used to measure the current between the electrodes while
their distance was controlled using the piezo element. Calibration
of inter-electrode displacements were done by means described
elsewhere\cite{calibration}.
\\
Typical traces are shown in Fig. \ref{Traces}. The conductance increases
exponentially as the electrodes are brought together, followed by a
plateau of conductance that reveals the formation of a one-atom contact.
In previous
works\cite{Review,First,Gabino,Cross,Sirvent,tungsten,Adatoms,Ugarte,Sirvent2}
this process has been studied experimentally mainly by looking at
individual traces of conductance. Although this provides some information
on the process, it is not suitable for quantitative conclusions. Here we
have used another approach: First, we take a conductance histogram
%(such as the one in the right panel of Fig. \ref{Density_Au})
from a few thousands of traces of conductance. The conductance histogram
is used to identify the conductance at which the contact is formed
(G$_c$). Next, for each contact closing trace we automatically record two
values, namely the value of conductance below G$_c$ at which the steepest
jump of conductance occurs (G$_a$) and the value of conductance
immediately following the jump (G$_b$). Finally, these values are used to
construct a density plot such as the one in the left panel in
Fig.~\ref{Density_Au}. The density of pairs (G$_a$,G$_b$) are plotted in
color coding with G$_a$ and G$_b$ along the horizontal and vertical axis,
respectively. The pairs (G$_a$,G$_b$) for which a JC occurs with maximal
probability are shown as a maximum in this density plot.
\\
From all metals studied, three characteristic behaviors have been
identified. Figure~\ref{Density_Au} shows our results for the case
of Au from an analysis of more than 300\,000 traces. First of all
we notice that there are no data near the line G$_a$=G$_b$,
showing that there is always a JC. We observe a first maximum at
the conductance (0.03,0.87) in units of G$_0$, which means that in
most of the cases the tunneling regime ends at a conductance of
0.03 G$_0$ and then the conductance jumps to a value of 0.87
G$_0$. The distribution for the jumps is quite broad for this
maximum, especially in G$_a$.  Further analysis shows that this
peak is most likely the superposition of one at (0.01,0.83) and
another at (0.05,0.94). We notice a second maximum at (0.09,1.6),
which corresponds to 7\% of the data, as calculated from the
distribution shown at the right panel of Fig.~\ref{Density_Au}.
\\
In the case of W our observations coincide with those of
Ref.~\onlinecite{tungsten}, i.e., the JC is nearly always absent,
as seen in the left panel of Fig.~\ref{Density_W_Ir}. The
evolution of the traces of conductance as the contact is stretched
is completely different from those for Au since there is not a
sharp transition between conductance plateaus. In the density plot
we observe most of the data concentrated near the line defined by
G$_b=$G$_a$. Although there are some abrupt transitions, these
show no reproducibility and are seen in the density plot as the
scarce data away from the G$_b=$G$_a$ line.
\\
Finally, in the case of Ir, shown in the right panel of
Fig.~\ref{Density_W_Ir} we find two maxima, one clearly shown at
(1.1,2.4) G$_0$ and another on the G$_a$=G$_b$ line. The first
maximum corresponds to a JC as observed for Au. The conductance at
which the contact is formed is more than twice that of Au because
of the higher valence of Ir with respect to Au\cite{Elke}. The
valence of the metal determines the number of electronic channels
through the contact, and each channel contributes a conductance
with a maximum of the quantum unit, G$_0$. Note that the most
probable conductance in the tunneling regime just before the Ir
contact is formed is higher than one quantum of conductance. The
second peak is located close to the line G$_b=$G$_a$ and shows the
21\% of cases when the JC is absent or is too small to be
observed, which is similar to the behavior we observed for W.
\\
These three examples are representative of what happens during
contact formation. We have analyzed other materials and summarize
our results in the following table:
\begin{center}
\begin{tabular}{|c|c|c|c|} \hline
{\bf Metal} & {\bf G$_c$}(G$_0$)& {\bf JC}&{\bf maxima (G$_a$,G$_b$)}(G$_0$) \\
\hline \hline Au&1.0&100\%&(0.01,0.87);(0.05,0.94);(0.09,1.6)
\\Pt&1.5&100\%&(0.13,1.4);(0.19,2.0);(0.35,3.1)\\Ag&1.0&100\%&(0.07,1.0);(0.2,1.6)
\\Ir& 2.0&79\% &(1.1,2.4)
\\Ni&1.3&75\%&(0.2,1.2)
\\W&1.0&0\%& \\
\hline
\end{tabular}
\end{center}
This table shows, for the different materials the conductance of
the first plateau, the percentage of cases having a JC, and the
positions of the maxima that can be clearly identified from the
density plots. Note that for Pt three peaks are observed, like for
Au, while for Ag only two were resolved.
\\
The process of JC is related to the mechanical stiffness of the
contact in competition with the attractive binding forces between
the atoms to be contacted. Atomistic simulations \cite{Sutton,
Smith, Landman} have shown that when surfaces are brought into
contact a jump occurs due to the adhesion forces between the
metallic layers. However, this jump depends on the stiffness of
the material and the geometry of the surfaces involved. For
example, we have performed simple static calculations using
embedded atom potentials for different surface geometries. The
surfaces are shifted rigidly each step and then relaxed using
conjugate gradient algorithms. From these calculations we observe
a clear jump between two perfect (100) surfaces as well as for an
atomically sharp tip approaching a perfect surface. However, the
transition is smooth when two tips approach each other.
%Consequently, a JC would be expected from the first two
%configurations while no JC would be observed in the last case.
\\
In the case of the nanocontacts described above, it is therefore expected
that the existence of a JC will depend on the shape of the surfaces before
contact. In order to understand this effect we have performed molecular
dynamics simulations of the elongation of nanowires of Au and W until
fracture. These fractured structures are then dynamically brought into
contact. The simulations consisted of 525 and 437 atoms for Au and W,
respectively, where three atomic layers at the top and bottom of the
simulation box were displaced every 1000 time steps by a fixed distance,
resulting in deformation rates characteristic of these types of
calculations \cite{Daneses}.  The simulation temperature was fixed to 4.2
K. An EAM potential was used for the case of Au \cite{Daw} while a
Finnis-Sinclair potential was used for W \cite{Finnis}. A total of 25
cases were computed both for Au and W. The atoms that first come into
contact were identified considering a cut-off distance between first and
second nearest neighbors. The distance between these atoms is then
calculated as the two sides approach. Fig.~\ref{Simulation01} shows the
distance between the pairs of atoms that first come into contact, for two
different configurations. Notice that in both cases there is a jump at
contact. This occurs for all simulations performed for Au. The insets in
Fig.~\ref{Simulation01} show the configurations at the point of contact
for these two cases. In the left panel the contact is made by two atoms,
one on each side, forming a dimer, while in the right panel a single atom
(monomer) from one side contacts three atoms on the opposite side. From
the 25 cases simulated 72\% dimers are formed at contact, 20\% monomers,
while 8\% are contacts through two parallel dimers or a double-bond. The
existence of three different configurations is in agreement with the
experimental observations for Au, where three peaks are observed in the
density plot. Moreover, comparing the statistics obtained from these
calculations with the experimental values we infer that the third peak for
Au in the table above corresponds to a double-bonded configuration, while
the first two peaks correspond to monomer and dimer configurations.
\\
In order to test the validity of the interpretations of the experimental
observations, the conductance was computed using the first-principles
transport code ALACANT\cite{alacant} for representative cases in all three
configurations mentioned above. We have used either local density or
generalized gradient corrected approximations for the exchange-correlation
potential and pseudopotential basis sets with system sizes of up to 100
atoms. Although a statistical analysis is not feasible, the results lie
within the experimental distribution and confirm the above established
correspondence between preferred conductance values and specific atomic
configurations, that is, the third peak for Au corresponds to a
double-bonded configuration, while the first two peaks would correspond to
monomer and dimer configurations respectively.
\\
 Finally, W was studied
by molecular dynamics in order to understand those cases where no JC
occurs. Figure~\ref{Simulation02} shows a particular configuration for W
at the point of contact. Unlike Au, for W there are configurations for
which a jump is not observed at contact. These configurations are closer
to two ideal tips than those found for Au. For a brittle material such as
W much sharper and ordered structures are expected than for more ductile
materials such as Au. In fact, as pointed out previously \cite{tungsten},
the elastic properties of these metals play an important role in the
presence of a JC.
\\
In conclusion, a systematic analysis of the occurrence of a jump to
contact in different metals has been performed using MCBJ and STM
experiments. These experiments demonstrate the absence of a jump to
contact for several metals. The results are interpreted using atomistic
calculations of Au and W. The geometry of the electrodes before contact
seem to play a very significant role determining whether a JC occurs, and
this shape is related to the elastic properties of the metals.
Consequently, softer metals such as Au or Ag always present a JC while
more brittle ones such as W do not. Moreover, with the help of
first-principles transport calculations we have been able to correlate the
preferential values of conductance measured at JC in Au with three
distinct configurations: a dimer, a monomer and a double bond.

This work is part of the research program of the ''Stichting FOM''.
Financial support through FIS2004-02356 and MAT2005-07369-C03-1 of the
Spanish MEC, and ACOMP06/138 of the Generalitat Valenciana is acknowledge.
C.\,U.\, and M.\,J.\,C.\, acknowledge a "Ram\'on y Cajal" grant of the
Spanish MCyT. We are grateful to David Jacob, S.\,Vieira  and his group,
and to J.\,L McDonald for technical, scientific and computer support.
%C.\,U.\,, M.\,J.\,C.\, and R.\,C.\, acknowledge support of the Spanish MEC
%under grant FIS2004-02356 and Generalitat Valenciana under grant
%ACOMP06/138. C.\,U.\, and M.\,J.\,C.\, acknowledge a "Ram\'on y Cajal"
%grant of the Spanish MCyT. J.J.P. acknowledges support from Spanish MEC
%under grant MAT2005-07369-C03-1 and technical support from David Jacob. We
%are grateful to the group of S.\, Vieira for their scientific and
%technical support and to J.L McDonald for computer support.
\\

\end{document}